\documentstyle[aps,prl,multicol,epsf,eqsecnum]{revtex}
\renewcommand{\v}[1]{{\bf #1}}
\renewcommand{\narrowtext}{\begin{multicols}{2} \global\columnwidth20.5pc}
\renewcommand{\widetext}{\end{multicols} \global\columnwidth42.5pc}

\begin{document}
\renewcommand{\v}[1]{{\bf #1}}
\renewcommand{\bar}[1] {{\overline #1}}
\newcommand{\be}{\begin{equation}}
\newcommand{\ee}{\end{equation}}
\newcommand{\ba}{\begin{eqnarray}}
\newcommand{\ea}{\end{eqnarray}}
\newcommand{\nn}{\nonumber \\}
\newcommand{\bbar}{b^{+}}
\newcommand{\cbar}{c^{+}}
\newcommand{\dbar}{d^{+}}
\newcommand{\fbar}{f^+}
\newcommand{\pbar}{p^+}
\newcommand{\wph}{\omega_{ph}}
\newcommand{\ed}{\epsilon_d}
\newcommand{\ek}{\epsilon_k}
\newcommand{\dtau}{\partial_\tau}
\newcommand{\dt}{\partial_t}
\newcommand{\dph}{D_{ph}}
\newcommand{\cL}{{\cal L}}
\newcommand{\mul}{\mu_L}
\newcommand{\mur}{\mu_R}
\newcommand{\dbarsigma}{d^+_\sigma}
\newcommand{\cbarksigma}{c^+_{k\sigma}}
\newcommand{\Gless}{G^<}
\newcommand{\Gmore}{G^>}
\newcommand{\Dless}{D^<}
\newcommand{\Dmore}{D^>}
\newcommand{\Gtbar}{G^{\overline{T}}}
\newcommand{\Dtbar}{D^{\overline{T}}}
\newcommand{\Xbar}{X^+}
\title{Keldysh Functional Integral Theory of Non-equilibrium Kondo Bridge}
\author{  Jung Hoon Han }
\address{Department of Physics, Sungkyunkwan University,
Suwon 440-746, Korea} \address{CSCMR, Seoul National University,
Seoul 151-747, Korea}
\date{\today}
\maketitle \draft

\begin{abstract}
We develop a consistent method for calculating non-equilibrium
Green's functions for a nano-sized dot coupled to electron reservoirs
by tunneling. The leads are generally at different chemical
potentials (non-equilibrium), and the dot may be a single molecule
characterized by its own vibrational frequency $\wph$. We carry out a
Keldysh functional analysis of the effective Green's functions of the
local electron in the dot, and also of the vibration mode, in the
Kondo regime. Finally, the tunneling current is calculated as a
function of bias using non-equilibrium Green's function. We find
stepwise increase of the current when bias exceeds even multiples of
the phonon frequency.
\end{abstract}

\bigskip
\narrowtext
\section{introduction}

Observing electron transport through nano-scale structures, often
consisting of a single molecule, is an active scientific endeavour
nowadays\cite{C60,molecular-kondo,recent-experiment,qd-kondo,kastner}.
With the ability to fabricate, and to control the conditions of, such
small-scale objects, we have the unique opportunity to explore the
subtle many-body phenomena under non-equilibrium conditions. A good
example of this will be the Kondo phenomenon occurring in a quantum
dot flanked by a current source and
drain\cite{molecular-kondo,recent-experiment,qd-kondo,kastner}.
Fixing the leads' chemical potentials at different values lead to
``non-equilibrium Kondo phenomena".

Unlike the solid-state quantum dot that has been studied extensively
in the past few years\cite{qd-kondo,kastner}, a molecular dot
contains a new degree of freedom associated with a conformational
change. Such mode can be typically described by an optical phonon of
frequency $\wph$. As a consequence, the tunneling current exhibits a
step-wise increase as the bias voltage passes through an integer
multiple of the vibration frequency $\wph$\cite{C60}. The coupling of
phonon mode to electron transport is an interesting feature absent in
a more conventional quantum dot.

It is conceivable that the same electron-phonon coupling takes place
in the Kondo regime, {\it i.e.} when the electrical transport is
carried out by the Abrikosov-Suhl resonance state formed across the
dot and the leads. A signature of phonon-assisted Kondo peak has been
reported in a recent experiment using $C_{60}$ molecule as a
tunneling bridge\cite{recent-experiment}. On the theoretical side,
Lee and Choi considered the modification of the Kondo temperature
scale $T_K$ when a phonon mode is present in the dot and found that,
due to the decreasing value of effective Hubbard parameter $U$, one
should expect increased $T_K$\cite{lee-choi}. Diagrammatic treatment
of the Kondo effect in the presence of a.c. perturbation and under
non-equilibrium is also present in the literature\cite{diagrams}.

Despite these early achievements, more progress can be made to
develop easy-to-use techniques for sorting out non-equilibrium
many-body phenomena. For example, the essence of Kondo resonance in
equilibrium can be obtained straightforwardly in the functional
integral formalism developed by Read and Newns (RN)\cite{read}. Here,
a slave-boson representation of the free energy of the problem is
expressed as an imaginary-time path-integral and subsequently the
stationary-phase condition is applied to look for the mean-field
state. The mean-field solution precisely corresponds to the
Abrikosov-Suhl resonance state. On the other hand, the first
pioneering study of the non-equilibrium Kondo physics in the Anderson
model taken up by Meir, Wingreen, and Lee used non-equilibrium
generalization of the non-crossing-approximation (NCA) and
equation-of-motion (EOM) techniques\cite{meir-wingreen}. Aguado and
Langreth recently introduced the non-equilibrium generalization of
the slave-boson theory of the Kondo resonance in their study of the
transport through double quantum dot while avoiding the direct use of
the free-energy minimization principle of RN\cite{langreth}.

In this paper, we take the first step towards a consistent
generalization of the functional integral theory of RN to the case of
non-equilibrium Kondo physics. Instead of the imaginary time
integral, we write down the real-time action over the Keldysh
contour. Integrating out the lead fermion degrees of freedom, the
effective action for the local fermion in the dot is derived that
agree with the expressions obtained by earlier
authors\cite{meir-wingreen,langreth}. The method is then applied to
study the phonon-coupled Kondo resonance under non-equilibrium. We
obtain the linear damping rate of the phonon mode due to coupling to
the current at finite bias $V$. Furthermore, current vs. voltage
formula is obtained for this case that exhibits the step-wise
increase in the current, {\it at even multiples of $\wph$}, for
strong electron-phonon coupling and for a narrow width of the
resonance spectrum.

Before we go into the details of the result it is worthwhile to note
a similar effort by Komnik and Gogolin, who combined effective action
approach with the Keldysh technique to work out the mean-field theory
of the Anderson impurity model under non-equilibrium
condition\cite{gogolin}.

The following sections are organized as follows. Section II
introduces the main formalism through discussion of the
non-equilibrium  Kondo resonance. Section III is the generalization
to phonon-coupled Kondo problem, in non-equilibrium. Current-voltage
expression for the phonon-coupled Kondo model is derived in section
IV. We close with a summary and outlook in section V.

\section{Non-equilibrium Kondo resonance}

We consider the simplest non-equilibrium case of two leads with
chemical potentials $\mul$ and $\mur$, connected by tunneling to a
quantum dot. We employ the model Hamiltonian $H = H_L + H_R + H_d +
H_T $ where \ba H_{L(R)} &=&\sum_{k\sigma} (\epsilon_k -\mu_{L(R)}
)\cbar_{L(R)k\sigma}c_{L(R)k\sigma} \nn H_d &=& -\ed  n_d + U n_d
(n_d -1) \nn H_T &= &V_L \sum_{k\sigma} \left( \cbar_{Lk\sigma}
d_\sigma + \dbarsigma c_{Lk\sigma}\right) + (L\rightarrow R),
\label{nonequil-H}\ea to study such system. We denote the local
fermion occupation by $n_d =\sum_\sigma \dbar_\sigma d_\sigma$. There
is no notion of well-defined set of eigenstates for a non-equilibrium
problem such as this one. Consequently, one is unable to write down
the free energy for the problem according to the textbook recipe.
This is why, we believe, naively writing down the imaginary-time
action for $\mul\neq\mur$ leads to incorrect results. Instead, we
write down the Lagrangian density in real-time coordinate along the
Keldysh contour $[t_i, t_f]\times [t_f, t_i]$ with $t_f - t_i = T$.
Eventually we take $t_i \rightarrow -\infty$, and $t_f \rightarrow
\infty$. First, the Lagrangian density for the Kondo problem, Eq.
(\ref{nonequil-H}), in the slave-boson representation ($U\rightarrow
\infty$) is written down:\widetext \ba \cL (t) & = & \sum_{k }
\cbar_{Lk\sigma}(i\dt -\ek+\mul) c_{Lk\sigma} - V_L \sum_{k}
[\cbar_{Lk\sigma} \bbar f_\sigma + \fbar_\sigma b c_{Lk\sigma} ] +
(L\rightarrow R) \nn &+& \fbar_\sigma (i\dt +\ed) f_\sigma + \bbar
i\dt b - \lambda ( \fbar_\sigma f_\sigma + \bbar b -1). \ea
\narrowtext\noindent (Through the rest of the paper we assume the
spin summation and drop $\sum_\sigma$.) This Lagrangian density is
integrated along the Keldysh contour $C$ to give the action $S=\int_C
dt \cL (t)$. From this we isolate the dynamics of the dot by
integrating out the lead electrons $c_{Lk\sigma}, c_{Rk\sigma}$. \ba
&&\cL_{eff}= - \fbar_\sigma b [V_L^2\sum_k G_{Lk} +V_R^2 \sum_k
G_{Rk} ] \bbar f_\sigma \nn &&~~~+ \fbar_\sigma (i\dt + \ed) f_\sigma
+ \bbar i\dt b \!-\! \lambda ( \fbar_\sigma f_\sigma \!+\! \bbar b
\!-\!1). \ea The self-energy part is defined over the double Keldysh
contour $\int_C \int_C dt dt'$ while the remaining terms are over a
single contour, $\int_C dt$. Green's functions for the lead electrons
are given by $G_{L(R)k} =(i\dt -\ek +\mu_{L(R)})^{-1}$. At this
stage, we take the mean-field ansatz, $b(t)=b$ and
$\lambda(t)=\lambda$, over the entire contour $C$. When this is done,
in fact the $\lambda (b^+ b -1)$ part drops out from the Lagrangian
because the forward and backward integrals cancel out. We thus have
the effective action ($G_L =\sum_k G_{Lk}, G_R =\sum_k G_{Rk}$,
$\lambda'=\lambda-\ed$, $V_{L(R)}' = b V_{L(R)}$) \ba &&
~~~~~~S_{eff} = \int_C dt \fbar_\sigma (t) (i\dt -\lambda')
f_\sigma(t) \nn &&- \int_C dt dt' \fbar_\sigma (t) [ V_L^{'2}
G_{L}(t,t') + (L\rightarrow R) ] f_\sigma (t'). \label{S_eff} \ea

The Green's function $G_{L(R)}(t,t')$ may be labeled
$(++),(+-),(-+),(--)$ depending on whether the time argument $t,t'$
are on the upper (+) or lower (-) branch of the contour. In turn, we
can associate each Green's function with $G^T$, $\Gless$, $\Gmore$
and $\Gtbar$ that are commonly used in the non-equilibrium theory as

\ba G^{++} (t-t') = G^T (t-t') &=& -i \langle T \psi(t)\bar \psi (t')
\rangle \nn G^{+-} (t-t')=\Gless (t-t') &=& i \langle \bar \psi (t')
\psi(t) \rangle \nn  G^{-+} (t-t')=\Gmore (t-t') &=& -i \langle
\psi(t)\bar \psi (t') \rangle \nn G^{--} (t-t')=\Gtbar (t-t') &=& -i
\langle \bar T \psi(t)\bar \psi(t') \rangle. \nonumber\ea In addition
retarded and advanced Green's function may be obtained readily by $
G^{R} = G^T  - \Gless $,  $G^A = G^T -\Gmore $.

We must also distinguish the fermion operators defined in the upper
and lower branches separately, as $F_\sigma = \left(
\begin{array}{c} f_{\sigma + } \\ f_{\sigma-} \end{array}\right)$.
In Fourier space, the effective action of Eq. (\ref{S_eff}) becomes
\[ \sum_\omega F^+_{\sigma }(\omega) \left(\begin{array}{cc} \omega
\!\!-\!\! \lambda'
\!\!-\!\!\Sigma^{++} & \Sigma^{ +-}  \\
\Sigma^{-+} & -\omega \!\!+\!\! \lambda' \!\!-\!\!\Sigma^{ --}
\end{array} \right)
F_{\sigma }(\omega), \] where we introduced the simplifying notation
$\Sigma^{ab} = (V_L^{'2} G_L^{ab} +V_R^{'2} G_R^{ab})$, $a,b=+,-$,
and $\sum_\omega =\int {d\omega \over 2\pi}$.

The inverse of the $2\times2$ matrix in the above effective action is
the Green's function for the local fermion. Lead Green's functions
are readily obtained \ba
\left(\begin{array}{cc}G_{L}^{T} & \Gless_{L}  \\
\Gmore_L & G_{L}^{\overline{T}} \end{array}\right) = -i\pi\rho
\left(\begin{array}{cc} {\rm sgn} (\omega-\mul) &
-2 \theta(\mul -\omega)  \\
2\theta(\omega-\mul ) &  {\rm sgn} (\omega-\mul)\end{array}\right),
\label{G_bath}\ea assuming a constant density of states $\rho$.
Similar definitions apply for the electrons in the right lead. We
assume that the left/right leads have the same density of states. On
inverting the matrix, we get for the local fermion Green's function
\newpage
\widetext \ba \v G_f &=&  \left(\begin{array}{cc} G_f^T & G_f^{<} \\
G_f^> & G_f^{\overline{T}} \end{array}\right) = {1\over
(\omega-\lambda')^2 \!+\! \Gamma^2 } \times \nn
&&\left(\begin{array}{cc} \omega \!\!-\!\! \lambda'
\!\!-\!\!i(\Gamma_L {\rm sgn} (\omega\!-\!\mul) +\Gamma_R {\rm sgn}
(\omega\!-\!\mur)) & 2i (\Gamma_L \theta (\mul \!-\!\omega) \!+\!
\Gamma_R
\theta(\mur \!-\!\omega))  \\
-2i (\Gamma_L \theta (\omega\!-\!\mul ) + \Gamma_R
\theta(\omega\!-\!\mur )) & -\omega\! +\! \lambda' \!\!-i(\Gamma_L
{\rm sgn} (\omega\!-\!\mul) \!+\!\Gamma_R {\rm sgn}
(\omega\!-\!\mur))
\end{array}\right),\label{G_f}\ea \narrowtext\noindent where
$\Gamma_L = \pi\rho V_L^{'2}$, $\Gamma_R = \pi\rho V_R^{'2}$ and
$\Gamma=\Gamma_L + \Gamma_R$. The non-equilibrium Green's functions
obtained in this way are identical to those obtained by earlier
authors\cite{meir-wingreen,langreth}. However, we argue that the
present method is much simpler in the sense that only the Gaussian
integration is involved in evaluating the effective Green's function.
One should introduce two species of fermions, for the upper/lower
branches of the contour, but can otherwise proceed with the usual
theory of integration over quantum fields. The retarded Green's
function is given by $ G_f^R = G_f^T - G_f^< = 1/( \omega-\lambda' +
i\Gamma )$. The spectral function, given as the imaginary part of
$G_f^R$, indicates the formation of a resonance state of width
$\Gamma$, centered at the energy $\lambda'$.

Determination of the precise value of $\lambda'$ is hampered by the
lack of precise definition of the free energy at hand, as discussed
at the beginning of this section\cite{coleman}. How to obtain a
suitable generalization of the free energy minimization principle for
non-equilibrium is outside the scope of the present paper. Instead,
we resort to ref. \cite{langreth} to conclude  $\lambda' \approx
(\mul + \mur )/2$. Therefore, a slave-boson theory in non-equilibrium
predicts a single Kondo resonance peak centered half-way between the
chemical potentials of the leads as noted earlier\cite{langreth}.

\section{Phonon-coupled Kondo resonance at non-equilibrium}

\begin{figure}
\hskip 0.9cm \epsfxsize=5.5cm\epsfysize=2.5cm \epsfbox{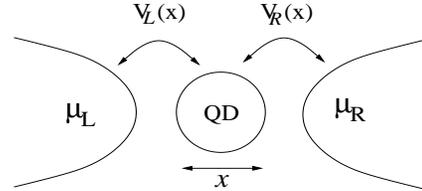}
\caption{Schematic setup of the phonon-coupled Kondo model.
Left/right leads of chemical potentials $\mul$ and $\mur$ are
connected to a quantum dot (QD) by tunneling. The tunneling
amplitudes $V_{L(R)}(x)$  depend on the center-of-mass coordinate $x$
of the QD.} \label{QD-setup}
\end{figure}

A molecular structure placed between the metallic leads, as well as
serving as a bridge for electrons to hop over, may vibrate thermally,
or quantum-mechanically in response to interaction with the passing
electrons. Such phonon signature in the $I-V$ characteristics has
been first detected in the tunneling experiment with the $C_{60}$
molecule. In this experiment, the observed phonon mode is that of the
center-of-mass displacement of the molecule as a bridge\cite{C60}. A
schematic drawing of the experimental setup is shown in Fig.
\ref{QD-setup}.

The geometrical distance between the molecule and the lead varies as
the center-of-mass position, denoted by $x$, swings back and forth.
The tunneling amplitude, being proportional to the overlap of the
wavefunctions in the lead and the molecule, becomes dependent on $x$.
The tunneling Hamiltonian will be accordingly modified: \be H_T = V_L
(x) \sum_{k\sigma} \cbar_{Lk\sigma} d_\sigma +V_R (x) \sum_{k\sigma}
\cbar_{Rk\sigma} d_\sigma+h.c. \ee The motion of the molecule itself
is modeled with a harmonic oscillator of frequency $\wph$. In this
section, we discuss how the phonon mode is damped due to interaction
with the electrons that form the resonance. The influence of phonons
on the electronic sector has been considered in detail in Ref.
\cite{lee-choi}. Hereafter we adopt the term ``phonon-Kondo model" to
describe the situation we are interested in.

The phonon-Kondo model may be written in real-time, \widetext \ba \cL
(t) & = & \sum_{k \sigma} \cbar_{Lk\sigma}(i\dt -\ek+\mul)
c_{Lk\sigma}-b V_L (x) \sum_{k\sigma} [\cbar_{Lk\sigma}  f_\sigma +
\fbar_\sigma c_{Lk\sigma} ] + (L\rightarrow R) \nn &&~~~+
\fbar_\sigma (i\dt -\lambda' ) f_\sigma +{1\over 2\wph} [(\dt x)^2 -
\wph^2 x^2]. \ea  We anticipate the mean-field condition
$b(t)=b,\lambda(t)=\lambda$ in writing down the above Lagrangian.
Getting rid of the lead fermions as before, the effective action
becomes \ba &&S_{eff} = - \int_C dt_1 dt_2 \fbar_\sigma (t_1 ) [ V'_L
(x(t_1 )) G_L (t_1 ,t_2 ) V'_L (x(t_2 )) ] f_\sigma (t_2 )
+(L\rightarrow R) \nn &&  ~~~~~+ \int_C dt\fbar_\sigma (t) ( i\dt
-\lambda' )f_\sigma (t)+{1\over 2\wph} \int_C dt [(\dt x)^2 - \wph^2
x^2]. \ea \narrowtext\noindent We have once again introduced the
simplifying notation $V'_{L(R)}(x)=bV_{L(R)}(x)$.

The two remaining dynamical modes in the above effective action are
the local fermions and the phonon. We now integrate out the local
fermions to derive the effective dynamics of the phonons, in part to
demonstrate the utility of the Keldysh technique in describing the
boson dynamics in non-equilibrium. We treat $V'_L (x) \approx V'_0
-V'_1 x$, $V'_R (x) \approx V'_0 + V'_1 x$, and expand the effective
action to quadratic order in $x$. In particular we have the bosonic
part of the action \be \cL_{eff}\approx {1\over 2\wph} [(\dt x)^2 -
\wph^2 x^2]+ 2iV_1^{'2} {\rm Tr} [ G_f x G_e x ].
\label{phonon-effective-action}\ee Here $G_f$ refers to the inverse
of $[ i\dt -\lambda' -V_0^{'2} G_e ]$, derived in Eq. (\ref{G_f}) of
the previous section, and $G_e = G_L + G_R$ (See Eq. (\ref{G_bath})).
The self-energy portion of Eq. (\ref{phonon-effective-action}) can be
written out also in the form

\be 2i V_1^{'2} \int_C dt dt' X^T (t)
\left(\begin{array}{cc} K^{++} & K^{+-}  \\
K^{-+}  & K^{--}
\end{array} \right)(t-t' ) X(t'),
\ee where we introduced $X= \left(\begin{array}{c} x_{+}  \\
x_{-}\end{array}\right)$, and

\ba K^{++}(t) &=& [ G_f^T (t) G_e^T (-t) +G_f^T (-t) G_e^T (t  ) ]/2
\nn K^{+-}(t) &=& -[\Gless_f (t)\Gmore_e (-t)+ \Gmore_f ( -t)\Gless_e
(t ) ]/2 \nn K^{-+} (t) &=& -[\Gmore_f (t) \Gless_e (-t)+\Gless_f
(-t) \Gmore_e (t)]/2 \nn K^{--}(t) &=& [\Gtbar (t) \Gtbar(-t) +\Gtbar
(-t) \Gtbar(t)]/2 . \label{K}\ea

A product of Green's functions has the following identities
\begin{eqnarray} G_1^T (t)G_2^T (-t) =\theta(t) \Gmore_1 (t) \Gless_2 (-t)
+\theta(-t) \Gless_1 (t) \Gmore_2 (-t) \nn \Gtbar_1 (t) \Gtbar_2 (-t)
= \theta(t) \Gless_1 (t)\Gmore_2 (-t) +\theta(-t) \Gmore_1 (t)
\Gless_2 (-t) \nonumber\end{eqnarray} that one can use to show, in
Fourier components, $ K^{++} = K^{--}= {1\over 4}(A+B+C+D)$ , $K^{+-}
= {1\over2}(A+B)$, $K^{-+} ={1\over2} (C+D)$  where $A=\sum_\omega
\Gless_f (\omega) \Gmore_e (\omega-\nu), B = \sum_\omega \Gmore_f
(\omega) \Gless_e (\omega+\nu), C=\sum_\omega \Gmore_f (\omega)
\Gless_e (\omega-\nu), D=\sum_\omega \Gless_f (\omega) \Gmore_e
(\omega+\nu)$. By using Eqs. (\ref{G_bath}) and (\ref{G_f}) one can
readily calculate $A,B,C$, and $D$ and all the kernels in Eq.
(\ref{K}).

On inserting the self-energy expression back into Eq.
(\ref{phonon-effective-action}) and inverting the matrix, a
remarkably simple expression of the phonon Green's function emerges:
\widetext\be {\bf D} (\nu) =\left(\begin{array}{cc} D^T
 & D^<  \nn D^>  & \Dtbar
\end{array} \right)= {1\over \left({\nu^2 -\wph^2 \over 2\wph
}\right)^2 + V_1^{'4} (K^{+-}-K^{-+})^2 } \left(\begin{array}{cc}
{\nu^2-\wph^2 \over 2\wph} - 2i V_1^{'2} K^{--} & -2i V_1^{'2} K^{+-} \\
-2i V_1^{'2} K^{-+} & {\wph^2 - \nu^2  \over 2\wph }- 2i V_1^{'2}
K^{++} \end{array} \right). \ee \narrowtext\noindent Retarded Green's
function for non-equilibrium is the difference \be D^R = D^T -\Dless
= {2\wph \over \nu^2 -\wph^2 +2i\wph V_1^{'2} (K^{-+}-K^{+-}) }, \ee
using the relation $2(K^{+-}-K^{--})=K^{+-}-K^{-+}$. Straightforward
calculation yields $K^{-+}-K^{+-} =$\ba &&
 \rho \left( \tan^{-1}\left({\mu_L \!\!+\!\!\nu\!\!-\!\!\lambda' \over
\Gamma }\right)\!\!-\!\! \tan^{-1}\left({\mu_L
\!\!-\!\!\nu-\!\!\lambda' \over \Gamma }\right)\right) \nn && +
\rho\left(\tan^{-1}\left({\mu_R \!\!+\!\!\nu-\!\!\lambda' \over
\Gamma }\right)\!\!-\!\! \tan^{-1}\left({\mu_R
\!\!-\nu\!\!-\!\!\lambda' \over \Gamma }\right) \right) \nn &&
~~~~\approx \left({2\rho\Gamma \over (\mu_L \!\!-\!\!\lambda')^2 +
\Gamma^2} \!\!+\!\!{2\rho\Gamma \over (\mu_R \!\!-\!\!\lambda')^2 +
\Gamma^2}\right)\nu\ea for a pair of chemical potentials $\mul$ and
$\mur$ of the leads and for low frequency $\nu$. Mean-field condition
sets $|\mu_L -\lambda'|\approx |\mu_R -\lambda'| \approx V/2$ without
the phonon coupling. We do not expect much modification of this
conclusion with the phonon coupling turned on, since the strength of
the electron-phonon coupling is significantly reduced to $V'_1 =bV_1
\ll V_1$ after the Kondo resonance is established. We then get the
linear damping coefficient for the phonon \be \gamma(V) =
 {4\rho\Gamma V_1^{'2} \over (V/2)^2 +\Gamma^2 }. \ee Interestingly,
we find the damping rate that is independent of the renormalization
factor $b^2$ at zero-bias, $\gamma (V=0) = (4/\pi) V_1^2 /(V_L^2 +
V_R^2)$. Finite $V$ surprisingly reduces the damping of the phonons.

\section{Non-equilibrium Tunneling Current}
A step-wise jump in the tunneling current when measured as a function
of the bias voltage was observed in the experiment using the C$_{60}$
molecule as a bridge connecting the leads\cite{C60}. The cause of the
jump is the opening up of an additional conducting channel as the
bias voltage exceeds integer multiples of the vibration frequency
$\wph$ of the intermediate molecule. While theory of the conductance
through a vibrating molecule has been worked out earlier by several
groups\cite{flensberg,millis,varma}, none addressed the issue using
the non-equilibrium Green's function technique advocated in this
paper, nor in the Kondo regime as we do now. We show here that the
pronounced features of the experiment can indeed be shown to follow
from our formalism.

To this end, we consider the non-equilibrium Anderson-Holstein model
defined by the Hamiltonian consisting of the left/right leads, at
different chemical potentials $\mul$ and $\mur$, the hybridization
between the dot and the lead with the tunneling element $V_L$ and
$V_R$, now taken as constants, and the dot+phonon Hamiltonian  given
by $ H = -\ed n_d +U n_d (n_d -1)+ \wph \bbar b + g\wph (n_d -1)
(\bbar + b)$.  A well-known Lang-Firsov canonical transformation of
the operators, $\hat{O}\rightarrow e^{\hat{S}} \hat{O} e^{-\hat{S}}$,
$\hat{S}=gn_d (\bbar -b)$, transforms $U$ into $U-2g^2 \wph$, and the
tunneling Hamiltonian becomes \be V_L \sum_{k} (\cbar_{Lk\sigma} X
d_\sigma + h.c.) + (L\rightarrow R) \label{tunneling-H}\ee where $X=
e^{g(b-\bbar)}=e^{-g^2/2}e^{-g\bbar}e^{gb}$. At the same time, the
local electron operator assumes a composite form $d_\sigma
\rightarrow X d_\sigma$. Original electron-phonon coupling term has
been removed in the transformed Hamiltonian. We further re-write the
$d_\sigma$ operator in the slave-boson form $\bbar f_\sigma$ assuming
that the effective Hubbard parameter $U-2g^2\wph$ is large enough to
prohibit double occupation of the dot.

In the Hamiltonian after the Lang-Firsov transformation,
electron-phonon coupling takes place through the $X$-operators
appearing in Eq. (\ref{tunneling-H}). When the phonon relaxation is
fast, {\it i.e.} any change in the phonon numbers caused during the
electron tunneling process is relaxed fast, phonon number
distribution will be given by the equilibrium value. In this case one
can replace $X \approx \langle X \rangle_{equil.} \approx e^{-g^2
/2}$. The last equality holds at zero temperature. Now we have
achieved complete separation of the phonon and the electron dynamics
in the transformed Hamiltonian and the electron-phonon coupling
provides only the renormalization of the tunneling amplitudes,
$V_{L(R)} \rightarrow e^{-g^2/2}V_{L(R)}$. This is not to say that
phonons have no effect on the dynamics of the local electrons.
Rather, the electron Green's function, being a product of the phonon
and the transformed local fermion Green's functions, will depend on
the dynamics of both quantities.

We proceed to calculate the Green's functions for each operator. The
Keldysh action one obtains from Lang-Firsov-transformed Hamiltonian
will consist of separate pieces for the phonon and the fermion
degrees of freedom according to the simplification of the previous
paragraph. Therefore, the local electron Green's function is
calculated as $-i \langle d_\sigma (t) \dbar_\sigma \rangle = -i b^2
\langle X(t) \Xbar \rangle \langle f_\sigma (t) \fbar_\sigma \rangle
= b^2 \langle X(t) \Xbar\rangle \Gmore_f (t)$, where $\Gmore_f$
refers to the Green's function of the Lang-Firsov-transformed
fermions. The correlation function for $X$ is obtained
straightforwardly from the harmonic oscillator algebra: $\langle
X(t)\Xbar\rangle = e^{g^2 [e^{-i\omega t}-1]}$. After Fourier
transform, a simple expression for $\Gmore_d$ emerges: $\Gmore_d
(\omega) = b^2 e^{-g^2} \sum_{n=0}^\infty {1\over n!} g^{2n} \Gmore_f
(\omega-n\wph)$. In a similar manner we find $\Gless_d (\omega) = b^2
e^{-g^2} \sum_{n=0}^\infty {1\over n!} g^{2n} \Gless_f
(\omega+n\wph)$.

In calculating the current through the dot, we invoke the
Meir-Wingreen formula\cite{meir-wingreen} that requires the knowledge
of the tunneling density of states $\rho_d (\omega)$. A short
manipulation gives $ \rho_d (\omega) = (i/2\pi) [ \Gmore_d (\omega)
-\Gless_d (\omega)]$. Current at zero-temperature is proportional to
the integral of $\rho_d (\omega)$ in the energy window $\mur\le
\omega \le \mul$, assuming $\mul-\mur = V>0$. We find, up to a
proportionality constant, the current \ba && 2 \tan^{-1}
\left({V\over 2\Gamma} \right)  + \sum_{n=1}^\infty {g^{2n}\over
n!}\theta(V\!\!-\!\!n\wph)\times \nn &&  ~~~~\left\{\tan^{-1}
\left({V\over 2\Gamma} \right) \!+\!\tan^{-1}
\left({V\!\!-\!\!2n\wph\over 2\Gamma} \right)\right\}. \ea In
obtaining this expression we used the mean-field result, $\lambda'
\approx (\mul+\mur)/2$, valid in the Kondo regime. The current
formula depends on the combined broadening parameter $\Gamma=\Gamma_L
+\Gamma_R$.

\begin{figure}
\hskip -0.5cm \epsfxsize=9cm\epsfysize=2.5cm \epsfbox{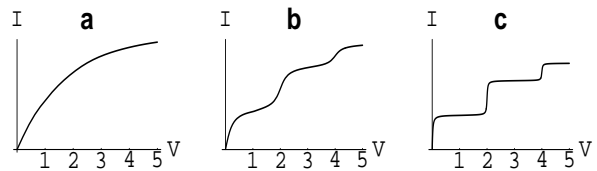}
\caption{Tunneling current vs. bias voltage (in units of $\wph$) for
several parameters $(g,\Gamma/\wph)=$ (a)(0.5,0.5), (b) (1,0.1) and
(c) (1,0.01). Steps are more pronounced for a smaller broadening
parameter $\Gamma$ and larger electron-phonon coupling $g$, and occur
at even multiples of $\wph$.} \label{IV-curves}
\end{figure}

Plot of the current vs. voltage for several values of broadening
parameter $\Gamma$ and electron-phonon coupling strength $g$ is shown
in Fig. \ref{IV-curves}. It shows that the step-wise behavior is most
pronounced for small values of $\Gamma/\wph$ and for large $g$. In
this regard, the Kondo resonance regime may be an excellent venue for
the observation of the phonon-assisted steps because of the small
$\Gamma$ value of the Abrikosov-Suhl state. Interestingly, the steps
occur at even multiples of the phonon frequency, and are absent for
odd multiples of $\wph$. This is to be distinguished from the steps
at both even/odd integer multiples of $\wph$ observed in $C_{60}$
experiment. The difference is due to the nature of the energy level
of the Kondo resonance state. In a resonant-tunneling model the
energy level of the dot is fixed regardless of the bias. In the Kondo
regime, however, the resonance is located at midway between the
chemical potentials of the leads, hence it ``floats" as the average
$(\mul+\mur)/2$ is varied. Effectively, the relative energy
difference between a lead and the dot level is half the overall bias
applied.
We have demonstrated that such non-equilibrium phenomena as the
phonon-assisted steps in the I-V characteristics can be consistently
derived using the formalism developed here.

\section{Summary and Outlook}

In this paper, we considered some theoretical aspects of transport
through a prototype single-molecule device, as experimentally
realized recently. A formalism has been developed that extrapolate
smoothly between equilbrium and non-equilibrium situations. In
summary, our method is based on writing down the real-time action
over the Keldysh contour and integrating out the lead electrons'
degree of freedom. This leaves us with the effective dynamics of the
localized electron in the dot, or for the phonon-coupled case, the
dynamics of the oscillatory motion of the dot. For example, we have
obtained the linear damping rate of the phonon mode in interaction
with the tunneling electrons in the Kondo regime. The tunneling
current through the phonon-coupled dot has been calculated as well;
it shows step-like increase in the current vs. voltage behavior.

Although the Kondo regime has been assumed through the discussion of
this paper, it is clear that our technique can be applied to any
out-of-equilibrium, resonant-level model. This work demonstrates the
utility and simplicity of the Keldysh functional integral approach
for dealing with non-equilibrium transport phenomena.

\section*{acknowledgment} The author thanks Tae-Suk Kim for bringing ref.
\cite{gogolin} to his attention, Jiwoong Park for helpful discussion
on the experiments, and Heung-Sun Sim for educating him on Kondo
physics. The author is supported by grant No. R01-2002-000-00326-0
from the Basic Research Program of the Korea Science \& Engineering
Foundation. Part of this work was carried out during participation at
the APCTP focus program on strongly correlated electrons.

\widetext

\begin{references}
\bibitem{C60} H. Park, J. Park, A. Lim, E. Anderson, A. Alivisatos,
and P. McEuen,  Nature (London) {\bf 407}, 57 (2000).

\bibitem{molecular-kondo} Jiwoong Park {\it et al.}, Nature (London) {\bf
417}, 722 (2002); Wenjie Liang {\it et al}, {\it ibid.} {\bf 417},
725 (2002).

\bibitem{recent-experiment} L. H. Yu and D. Natelson, NanoLetters {\bf 4},
79-83 (2004).

\bibitem{qd-kondo} D. Goldhaber-Gordon {\it et al.} Nature (London)
{\bf 391}, 156 (1998); W. G. van der Wiel {\it et al.} Science {\bf
289}, 2105 (2000).

\bibitem{kastner} Andrei Kogan, Sami Amasha, and M. A. Kastner, Preprint.


\bibitem{lee-choi} Hyun C. Lee and Han-Yong Choi,
Phys. Rev. B {\bf 69}, 075109 (2004).

\bibitem{diagrams} A. Kaminski, Yu V. Nazarov, and L. I. Glazman,
Phys. Rev. Lett. {\bf 83}, 384 (1999); A. Rosch, J. Kroha, and P.
W\"olfle, Phys. Rev. Lett. {\bf 87}, 156802 (2001).

\bibitem{read} N. Read and D. M. Newns, J. Phys. C {\bf 16}, L1055
(1983); N. Read and D. M. Newns, {\it ibid.} {\bf 16}, 3273 (1983).

\bibitem{meir-wingreen} Y. Meir, Ned S. Wingreen, and Patrick A.
Lee, Phys. Rev. Lett. {\bf 70}, 2601 (1993); Ned S. Wingreen and
Yigal Meir, Phys. Rev. B {\bf 49}, 11040 (1994).

\bibitem{langreth} Ram\'on Aguado and David C. Langreth, Phys. Rev.
Lett. {\bf 85}, 1946 (2000); Ram\'on Aguado and David C. Langreth,
Phys. Rev. B {\bf 67}, 245307 (2003).

\bibitem{gogolin} A. Komnik and A. O. Gogolin, Phys. Rev. B
{\bf 69}, 153102 (2004).

\bibitem{coleman} An effort to produce a sensible definition of the
free energy for non-equilibrium can be found in, {\it e.g.} P.
Coleman and W. Mao, J. Phys. Condens. Matter {\bf 16}, L263 (2004).

\bibitem{flensberg} Karsten Flensberg,
Phys. Rev. B {\bf 68}, 205323 (2003); Stephan Braig and Karsten
Flensberg, {\it ibid.} {\bf 68}, 205324 (2003).

\bibitem{millis} A. Mitra, I. Aleiner, and A. J. Millis,
cond-mat/0302132.

\bibitem{varma} Vivek Aji, Joel E. Moore, and C. M. Varma,
cond-mat/0302222.



\end{references}
\end{document}